\documentclass[aps,prl,twocolumn,preprintnumbers,superscriptaddress,amsmath,amssymb]{revtex4-1}

\usepackage{graphicx}
\usepackage{subfigure}
\usepackage{epsfig}
\usepackage{xcolor}
\usepackage{dcolumn}
\usepackage{bm}
\usepackage{ulem}
\usepackage{color}
\usepackage{amsmath}
\usepackage{amsfonts}
\usepackage{amssymb}

\usepackage{braket}

\begin{document}

\title{Experimental evidence of chiral symmetry breaking in Kekul\'e-ordered graphene}

\author{Changhua Bao}
\altaffiliation{These authors contributed equally to this work.}
\affiliation{State Key Laboratory of Low-Dimensional Quantum Physics and Department of Physics, Tsinghua University, Beijing 100084, P. R. China}

\author{Hongyun Zhang}
\altaffiliation{These authors contributed equally to this work.}
\affiliation{State Key Laboratory of Low-Dimensional Quantum Physics and Department of Physics, Tsinghua University, Beijing 100084, P. R. China}

\author{Teng Zhang}
\affiliation{School of Information and Electronics, MIIT Key Laboratory for Low-Dimensional Quantum Structure and Devices, Beijing Institute of Technology, Beijing 100081, P. R. China}

\author{Xi Wu}
\affiliation{Shenzhen Geim Graphene Center and Institute of Materials Research, Tsinghua Shenzhen International Graduate School, Tsinghua University, Shenzhen 518055, P. R. China}

\author{Laipeng Luo}
\affiliation{State Key Laboratory of Low-Dimensional Quantum Physics and Department of Physics, Tsinghua University, Beijing 100084, P. R. China}

\author{Shaohua Zhou}
\affiliation{State Key Laboratory of Low-Dimensional Quantum Physics and Department of Physics, Tsinghua University, Beijing 100084, P. R. China}

\author{Qian Li}
\affiliation{State Key Laboratory of Low-Dimensional Quantum Physics and Department of Physics, Tsinghua University, Beijing 100084, P. R. China}

\author{Yanhui Hou}
\affiliation{School of Information and Electronics, MIIT Key Laboratory for Low-Dimensional Quantum Structure and Devices, Beijing Institute of Technology, Beijing 100081, P. R. China}

\author{Wei Yao}
\affiliation{State Key Laboratory of Low-Dimensional Quantum Physics and Department of Physics, Tsinghua University, Beijing 100084, P. R. China}

\author{Liwei Liu}
\affiliation{School of Information and Electronics, MIIT Key Laboratory for Low-Dimensional Quantum Structure and Devices, Beijing Institute of Technology, Beijing 100081, P. R. China}

\author{Pu Yu}
\affiliation{State Key Laboratory of Low-Dimensional Quantum Physics and Department of Physics, Tsinghua University, Beijing 100084, P. R. China}
\affiliation{Frontier Science Center for Quantum Information, Beijing 100084, P. R. China}

\author{Jia Li}
\affiliation{Shenzhen Geim Graphene Center and Institute of Materials Research, Tsinghua Shenzhen International Graduate School, Tsinghua University, Shenzhen 518055, P. R. China}

\author{Wenhui Duan}
\affiliation{State Key Laboratory of Low-Dimensional Quantum Physics and Department of Physics, Tsinghua University, Beijing 100084, P. R. China}
\affiliation{Frontier Science Center for Quantum Information, Beijing 100084, P. R. China}

\author{Hong Yao}
\affiliation{Institute for Advanced Study, Tsinghua University, Beijing 100084, P. R. China}
\affiliation{Department of Physics, Stanford University, Stanford, California 94305, USA}

\author{Yeliang Wang}
\affiliation{School of Information and Electronics, MIIT Key Laboratory for Low-Dimensional Quantum Structure and Devices, Beijing Institute of Technology, Beijing 100081, P. R. China}

\author{Shuyun Zhou}
\altaffiliation{Correspondence should be sent to syzhou@mail.tsinghua.edu.cn}
\affiliation{State Key Laboratory of Low-Dimensional Quantum Physics and Department of Physics, Tsinghua University, Beijing 100084, P. R. China}
\affiliation{Frontier Science Center for Quantum Information, Beijing 100084, P. R. China}
\affiliation{Beijing Advanced Innovation Center for Future Chip, Beijing 100084, P. R. China}

\date{\today}

\begin{abstract}

{\bf The low-energy excitations of graphene are relativistic massless Dirac fermions with opposite chiralities at valleys K and K$^\prime$. Breaking the chiral symmetry could lead to gap opening in analogy to dynamical mass generation in particle physics. Here we report direct experimental evidences of chiral symmetry breaking (CSB) from both microscopic and spectroscopic measurements in a Li-intercalated graphene. The CSB is evidenced by gap opening at the Dirac point, Kekul\'e-O type modulation and chirality mixing near the gap edge. Our work opens up opportunities for investigating CSB related physics in a Kekul\'e-ordered graphene. }

\end{abstract}

\maketitle

\begin{figure*}[htbp]
	\centering
	\includegraphics[width=16.8 cm]{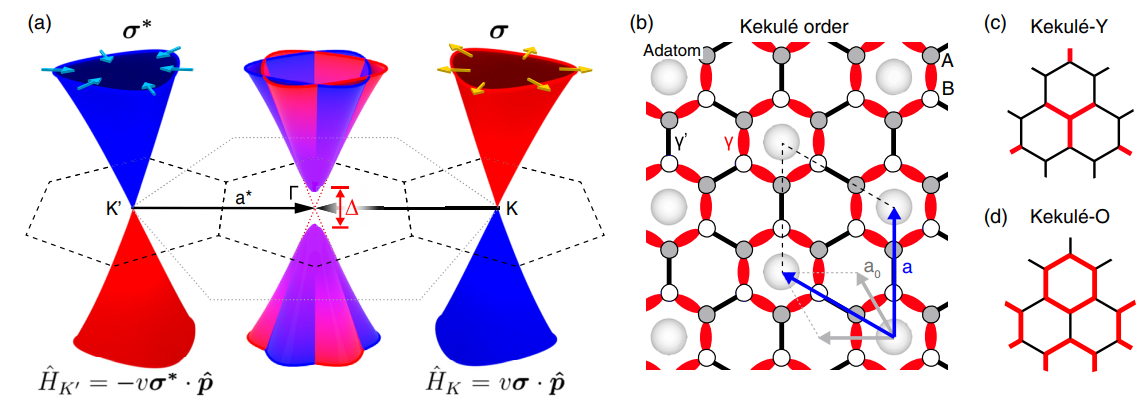}
	\caption{(a) A schematic for CSB. Straight blue and yellow arrows point to the pseudospin directions.  (b) A Kekul\'e-ordered graphene with different bond strengths (represented by black and red colors) and nearest-neighbor hopping parameters $\gamma$ and $\gamma^\prime$. The unit cells of pristine and Kekul\'e-ordered graphene are marked by gray and blue parallelograms. (c) Kekul\'e-Y bond texture. (d) Kekul\'e-O bond texture.}
\end{figure*}

Chirality $\hat{h}$ is a fundamental property for relativistic massless Dirac fermions,  which is defined by the projection of spin $\bm{\sigma}$ onto momentum $\bm{\hat{p}}$, namely, $\hat{h}=\frac{1}{2} \bm{\sigma} \cdot \bm{\hat{p}}/|\bm{\hat{p}}|$.  Chiral symmetry breaking (CSB), namely, coupling of Dirac fermions with opposite chiralities, leads to dynamical mass generation for elementary particles \cite{Nambu1961}, which lays a cornerstone for the Standard Model in particle physics. The low-energy excitations of graphene are massless Dirac fermions \cite{GRevRMP09,NetoRMP12} with the electron spin replaced by the pseudospin, and therefore graphene provides a condensed matter physics analogue for investigating CSB and mass generation indicated by band gap opening \cite{ChamonPRB2000,Roy2009,AltSchulerSSC,Seo2011}.
In addition to gap opening, a series of intriguing phenomena have been proposed with CSB, such as electron fractionalization \cite{MudryPRL2007} and topological effects in the electron and phonon spectra \cite{Hu2016,DuanKekule}.
Experimental realization and unambiguous observation of CSB in graphene are therefore highly desirable.

In order to realize CSB, a superlattice period of $(\sqrt3\times\sqrt3)R30^\circ$ is required to couple Dirac cones from K and K$^\prime$ valleys, and such inter-valley coupling could lead to replica Dirac cones at the Brillouin zone (BZ) center and gap opening as schematically illustrated in Fig.~1(a).  Such superlattice period is inherent in the Kekul\'e distortion with modulated bond structure as indicated by red and black bonds in  Fig.~1(b).
Experimentally, Kekul\'e distortion can be induced by a strong magnetic field \cite{LeeDH2009,Mudry2010,HeLPRB2019}, which however also breaks the time reversal symmetry and turns the conical dispersion into Landau levels.  Introducing dilute adatoms or defects \cite{AltSchulerSSC,PasupathyNatPhys, BeenakkerNJP, HeLACS2018} or an external superlattice potential \cite{MolecularGNat2012,Ortix2015,Zhang2017} provides an attractive pathway for realizing CSB while still preserving the conical dispersion.
Along this path, important progress has been made in graphene grown on Cu substrate,  where Kekul\'e-Y order with ``Y"-shaped bond modulation as shown in Fig.~1(c) has been revealed by scanning tunneling microscopy (STM) \cite{PasupathyNatPhys}. However, there is no evidence of gap opening, and recent theoretical calculation suggests that such Kekul\'e-Y ordered graphene would remain gapless, while Kekul\'e-O order where modulated bonds form an ``O''-shape pattern as shown in Fig.~1(d) is gapped \cite{BeenakkerNJP}.
Superlattice period of $(\sqrt3\times\sqrt3)R30^\circ$ has also been reported in Li or Ca intercalated graphene \cite{JohanssonARPES2010,TakahashiAIP2011,Takahashi2012,HitosugiSTMPRL2015,Jacobi2016,HeunBLtoG,HasegawaAPL2017} or graphite \cite{CaC6NP,MauriNatPhys2012LiC}, making them potential candidates for realizing CSB. However, experimental evidences such as CSB induced gap opening, specific type of Kekul\'e order and chirality mixing near the gap edge,  which are crucial for establishing the CSB, are still missing.

\begin{figure*}[htbp]
	\centering
	\includegraphics[width=16.8 cm]{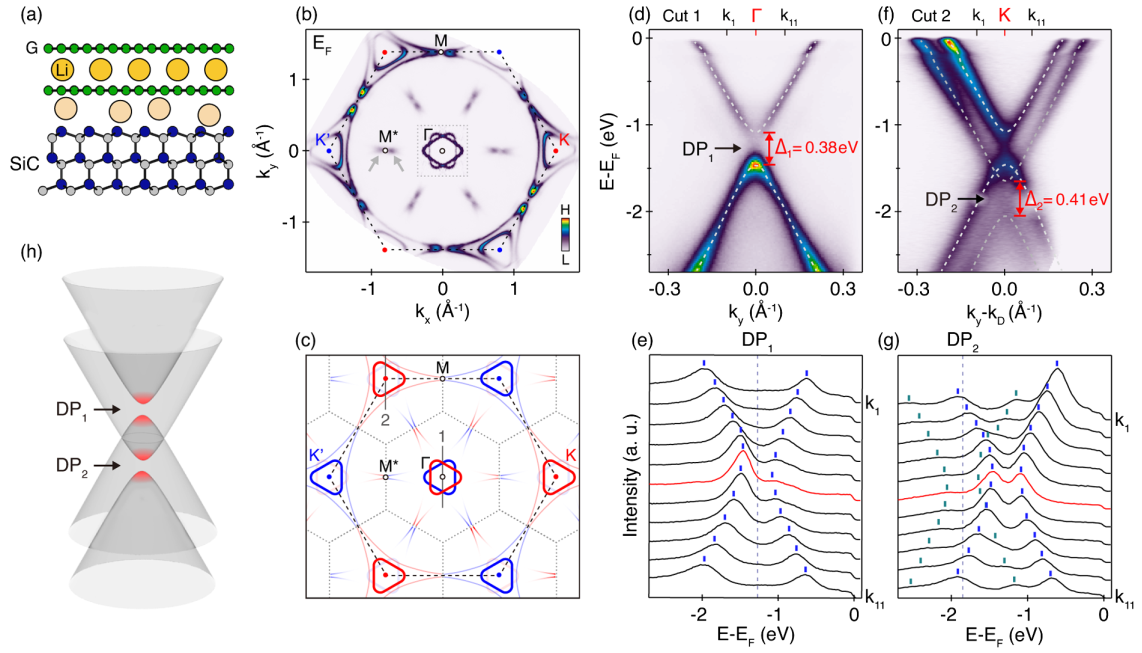}
	\caption{(a) A schematic drawing of  Li-intercalated double layer graphene on SiC. (b) Experimental Fermi surface map of the Kekul\'e-ordered graphene measured using helium lamp source at 21.2 eV.  The intensity inside the dashed box around $\Gamma$ is enhanced for better visualization.  (c) A schematic Fermi surface map of the Kekul\'e-ordered graphene. (d) Dispersion image of the folded Dirac cones measured along the M-$\Gamma$-M direction (marked by gray line in (c)). (e) EDCs for data shown in (d). (f) Dispersion image of Dirac cones through the K point and perpendicular to K-M direction (marked by gray line in (c)). (g) EDCs for data shown in (f). (h) A schematic summary of two gapped Dirac cones as observed experimentally. }
\end{figure*}

Here by combining  angle-resolved photoemission spectroscopy (ARPES) and STM measurements, we provide direct experimental evidences for CSB in a Li-intercalated graphene from both electron spectroscopic and microscopic measurements.  The CSB is confirmed by CSB induced gap opening in the Dirac cone, Kekul\'e-O type texture in the surface topography and chirality mixing near the gap edge.

The Kekul\'e order is introduced by intercalating Li \cite{JohanssonARPES2010} to monolayer graphene on SiC substrate (see Fig.~S1 and Supplemental Material for more details \cite{supp}).
The intercalation releases the bonding between the buffer layer \cite{Magaud2007} and the SiC substrate, which transforms the buffer layer into a new graphene layer \cite{JohanssonARPES2010,HeunBLtoG} and results in two graphene layers in the AA stacking \cite{HasegawaAPL2017}  as schematically illustrated in Fig.~2(a).  We note that intercalation of the buffer layer does not lead to Kekul\'e order \cite{Johansson2010,HeunBLtoG}, and two graphene layer structure as shown in Fig.~2(a) has the minimum thickness required to stabilize the Kekul\'e order. After intercalation, the Fermi surface map in Fig.~2(b) shows two trigonal pockets at each BZ corner, a hexagonal star-shaped pocket at the $\Gamma$ point and arcs near the M$^*$ point.
The small and large triangular pockets originate from the top and bottom graphene layers respectively, with carrier concentrations of 1.3$\times$10$^{14}$ cm$^{-2}$ and 4.3$\times$10$^{14}$ cm$^{-2}$ calculated from the size of the Fermi pockets using the Luttinger theorem \cite{Luttinger1960}. The star-shaped pocket at $\Gamma$ is formed by the superposition of  the folded inner triangular pockets from K and K$^\prime$ by the reciprocal  Kekul\'e superlattice vector (see schematic drawing in Fig.~2(c)).   Replicas of the larger triangular pockets folded from K and K$^\prime$ to $\Gamma$, and K to K$^\prime$ or vice versa are also observed near the M$^*$ point (pointed by gray arrows in Fig.~2(b)). The existence of folded Dirac cones by $(\sqrt3\times\sqrt3)R30^\circ$ superlattice modulation \cite{TakahashiAIP2011,Takahashi2012,Damascelli2015,MauriNatPhys2012LiC}  is not sufficient for generating a mass gap, and further experimental evidences are needed to confirm the CSB with a gap opening, which is the main focus of this work.

An important evidence is the gap opening at the Dirac point, which is revealed by ARPES dispersions measured through both $\Gamma$ and K points shown in Fig.~2, and more data around these points are shown in Figs.~S2 and S3 \cite{supp}.
A gap of 380 $\pm$ 10 meV is clearly identified  for the upper Dirac cone (inner pocket) from the dispersion image shown in Fig.~2(d) as well as energy distribution curves (EDCs) shown in Fig.~2(e).  Such gap value corresponds to an effective mass of $m^*$ = 0.033 $m_e$, where $m_e$ is the free electron mass.
The gap opening is also observed in the dispersion image measured along an equivalent cut through the K point (Fig.~2(f)).  Here, not only the upper Dirac cone but also the lower Dirac cone with a larger pocket is clearly resolved. The upper Dirac cone shows a gap value consistent with Fig.~2(d) with the Dirac point at -1.27 eV (broken line in Fig.~2(e)), and the lower Dirac cone shows a gap of 410 $\pm$ 50 meV ($m^*$ = 0.036 $m_e$) with the Dirac point at -1.82 eV (broken line in Fig.~2(g)).  Our ARPES data show that these two Dirac cones are both gapped and shifted in energy, as schematically summarized in Fig.~2(h). A gap was observed previously, yet its origin was elusive \cite{TakahashiAIP2011}. Here first-principles calculations are performed to support the CSB origin of the gap. The calculated band dispersion (Fig.~S5(a),(d) \cite{supp}) with the most stable Li configuration (Fig.~S4 \cite{supp}) is in agreement with the experimental observation. Moreover, a comparison of calculated dispersions with either the bottom or top Li layer removed (Fig.~S5 \cite{supp}) shows that the intercalated Li layer between the two graphene layers contributes dominantly to the gap opening. Because this Li layer is also critical for the formation of the $(\sqrt3\times\sqrt3)R30^\circ$ Kekul\'e order, our comparison indicates a close correlation between the gap opening and the Kekul\'e order.

\begin{figure*}[htbp]
	\centering
	\includegraphics[width=16.8 cm]{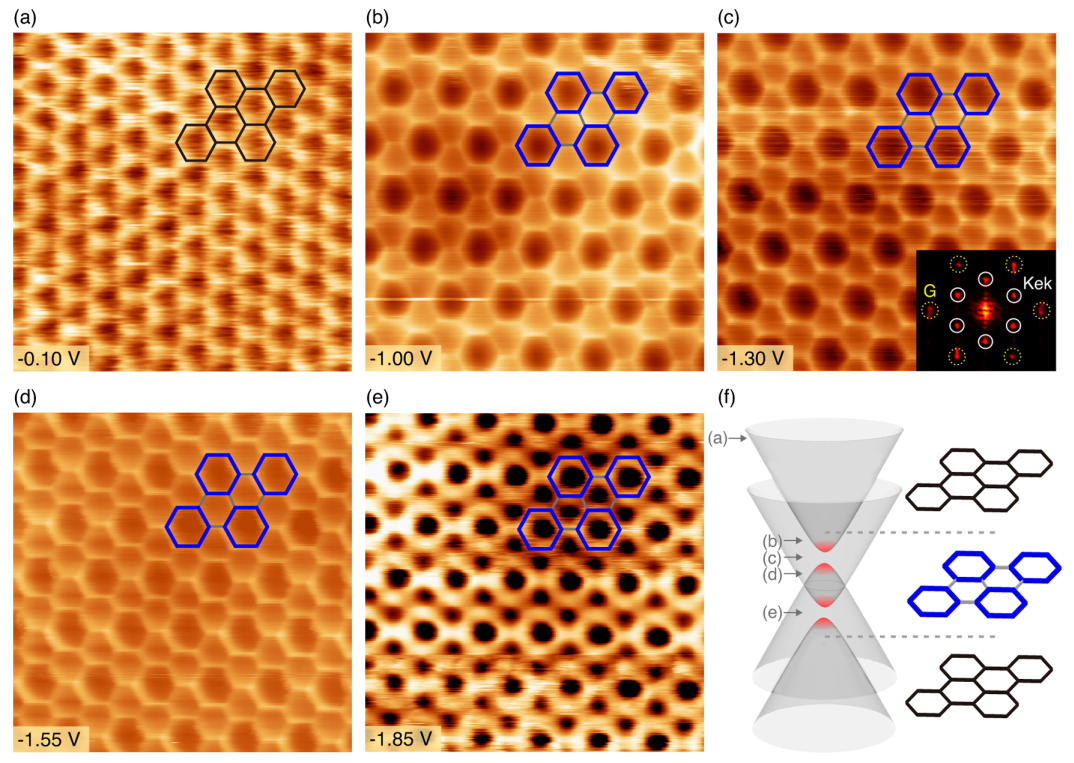}
	\caption{(a)-(e) STM topographic images at different bias voltages. The size of all images is  2.5 nm$\times$2.5 nm and the tunneling current is 0.5 nA (a) and 4.1 nA ((b)-(e)). The inset in (c) is two-dimensional fast Fourier transformation image of data in (c). (f) A schematic summary of Kekul\'e-O type modulation which is observed only near the gap edge.}
\end{figure*}

As noted above, recent theoretical calculation shows that only a specific type of Kekul\'e order can generate a gap opening \cite{BeenakkerNJP}, and therefore it is important to identify the specific type of Kekul\'e order. Here we directly probe the spatially-resolved electronic properties by performing STM topographic measurements at different bias voltages.  Away from the Dirac point energy, for example at -0.1 eV (Fig.~3(a)), the STM topography shows a honeycomb structure which is typical of graphene, while at energies near the gap regions of the two Dirac cones (Fig.~3(b)-(e)), the STM topography clearly demonstrates strongly modified topography, with a periodic arrangement of expanded hexagons (labeled by thicker blue bonds) alternating with highly distorted hexagons compatible with the Kekul\'e-O type pattern. Such Kekul\'e-O order is critical for inducing CSB gap \cite{BeenakkerNJP}. Compared to other decorated graphene \cite{Takahashi2012,HitosugiSTMPRL2015,PasupathyNatPhys,HeunBLtoG} where $(\sqrt{3}\times\sqrt{3})R30^\circ$ Kekul\'e order is also observed in the fast Fourier transformation image (inset of Fig.~3(c)), our work is the first to demonstrate Kekul\'e-O type modulation and to show that it exists only near the gap edge of the Dirac cone as schematically summarized in Fig.~3(f), by combining the advantages of both STM and ARPES measurements.
The fact that the Kekul\'e-O pattern is observed only near the gap edge indicates that it is a modulation of the electronic states rather than a simple modulation of the lattice which is energy independent. This suggests that although the CSB has a somewhat different origin (induced by an external superlattice potential imposed by the intercalated Li atoms rather than a spontaneous Coulomb interaction in quantum electrodynamics \cite{PisarskiPRD}), the results of the CSB, including CSB induced gap opening and modulated electronic states near the gap edge, remain the same.

\begin{figure*}[htbp]
	\centering
	\includegraphics[width=16.8 cm]{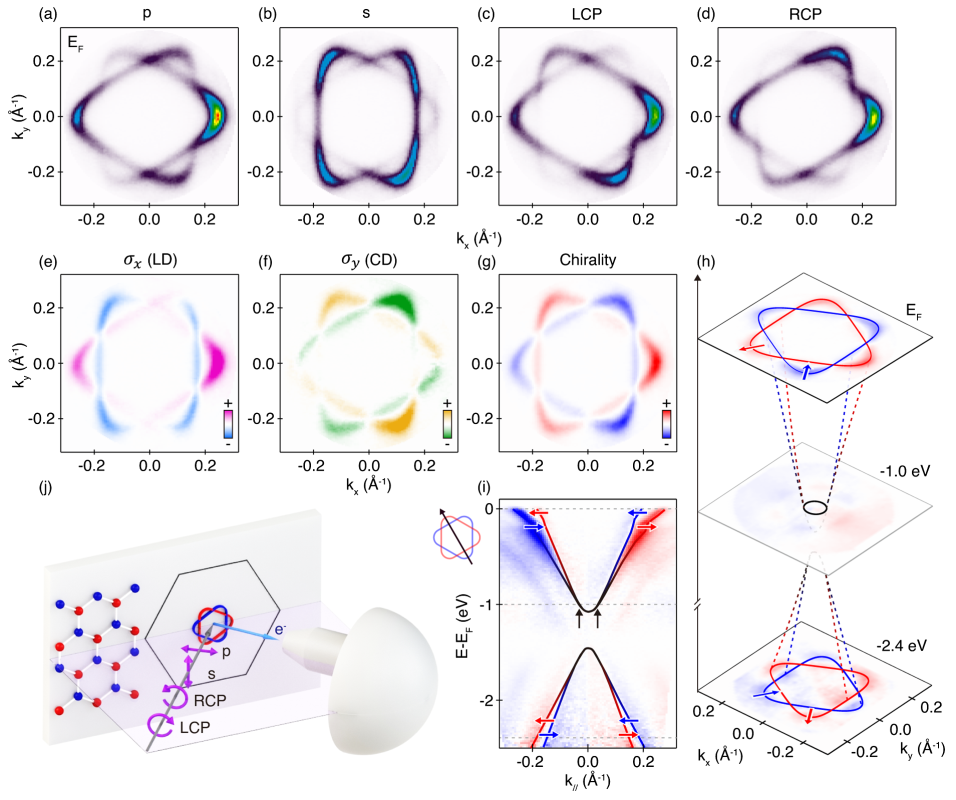}
	\caption{(a)-(d) Experimental Fermi surface maps measured around the $\Gamma$ point using 11 eV laser source with four different light polarizations. (e, f) Extracted $\sigma_x$ and $\sigma_y$ from data in (a)-(d). (g) Extracted  chirality maps at $E_F$. (h) Extracted  chirality maps at  $E_F$, -1.0 and -2.4 eV. Red, blue and black curves are guides for the pockets, and red and blue arrows indicate the pseudospin directions. (i) Chirality-resolved dispersion image.  The schematic cartoon on the left indicates the momentum direction where the dispersion is measured.  Black arrows point to chirality mixing near the gap edge. (j) Experimental geometry for the polarization-dependent ARPES measurements.}
\end{figure*}

We make one step further by pursuing the evolution of the chirality mixing near the gap edge through polarization-dependent ARPES measurements. Figure 4(a)-(d) shows Fermi surface maps measured using four different light polarizations, where the zigzag ($\Gamma$-K) direction is in the scattering plane (see experimental geometry in Fig.~4(j)).  The pseudospin selection rules can be derived following previous work \cite{ChiangCDARPES}, and similar measurements have been applied to reveal the Berry phase of graphene \cite{RotenbergPRB2008,LanzaraCDARPES,ChiangCDARPES}, which is the accumulated phase of the pseudospin around a closed loop.  Here we extend the analysis to further extract energy- and momentum-resolved pseudospin and chirality in our Kekul\'e-ordered graphene  (see Supplemental Material for more details on the analysis method \cite{supp}). In particular, the $x$ and $y$ components of the pseudospin for each point in the Dirac cone can be extracted from the linear dichroism (LD)  and circular dichroism (CD) ARPES by
$$I_{LD}=\frac{I_p-I_s}{I_p+I_s}=\begin{cases} \cos\theta = \sigma_x~(for~K~valley)\\  \cos\theta =\sigma^*_x~(for~K^{\prime}~valley)\end{cases}$$
$$I_{CD}=\frac{I_{LCP}-I_{RCP}}{I_{LCP}+I_{RCP}}=  \begin{cases} \sin\theta = \sigma_y~(for~K~valley)\\  -\sin\theta = \sigma^*_y~(for~K^{\prime}~valley)\end{cases} $$ where $I_p$, $I_s$, $I_{LCP}$ and $I_{RCP}$ correspond to ARPES intensity with two linear and two circular light polarizations.
Note that K and K$^\prime$ are related by the time reversal symmetry, $\hat{H}_{K^\prime}=-v \bm{\sigma^*}\cdot\bm{\hat{p}}$, and the pseudospin at K$^\prime$ is defined as $\bm{\sigma^*}=(\sigma_x,-\sigma_y)$ \cite{GRevRMP09}.

The  extracted momentum-resolved $\sigma_x$ and $\sigma_y$ maps from Fig.~4(a)-(d) are shown in Fig.~4(e),(f). By further taking the projection $\hat{h}=\frac{1}{2} \bm{\sigma} \cdot \bm{\hat{p}}/|\bm{\hat{p}}|$, opposite chiralities  (represented by red and blue colors) are observed for folded Dirac cones from K and K$^{\prime}$ at $E_F$ (Fig.~4(g)), endorsing the validity of this analysis method.
Moreover, chirality maps measured at different energies (Fig.~4(h)) and chirality-resolved dispersion image along the $\Gamma$-K direction (Fig.~4(i)) show that although a strong chirality contrast is observed away from the Dirac point (e.g., at $E_F$ and -2.4 eV), the chirality contrast becomes negligible in regions where folded Dirac cones from K and K$^\prime$ (blue and red curves) overlap (pointed by black arrows in Fig.~4(i)), indicating that there is no well-defined chirality near the gap edge of the upper Dirac cone.  We note that weak yet detectable chirality contrasts are observed at -1.0 eV on both sides of the upper Dirac cone, which are contributed by the lower Dirac cone away from the Dirac point (see Fig.~S6 \cite{supp}). Analysis of the energy-dependent chirality for the upper Dirac cone is shown in  Fig.~S7 \cite{supp}, which shows that the chirality contrast decreases gradually when moving toward the Dirac point and becomes negligible near the gap edge.
Such energy dependent chirality made available by polarization-dependent ARPES measurements confirms the chirality mixing near the gap edge, providing another important experimental evidence for CSB.

The observation of replica Dirac cones with CSB induced gap opening, Kekul\'e-O patterned topography and chirality mixing near the gap edge together provides definitive experimental evidences for CSB in a condensed matter physics system, in analogy to the mass generation in particle physics. Moreover, the experimental realization of CSB in the Kekul\'e-ordered graphene provides new opportunities for exploring CSB related physics, such as electron fractionalization \cite{MudryPRL2007} and topological effect \cite{Hu2016,DuanKekule}.

\begin{acknowledgments}
\section*{ACKNOWLEDGMENTS}
We thank D.-H. Lee, Y.-T. Huang, H. Ding, and T. Qian for helpful discussions. This work was supported by the
National Key R$\&$D Program of China (Grant No. 2016YFA0301004, 2020YFA0308800, 2016YFA0301001), the National Natural Science Foundation of China (Grants No. 11725418, No. 11427903), Beijing Advanced Innovation Center for Future Chip (ICFC), Tsinghua University Initiative Scientific Research Program, and Tohoku-Tsinghua Collaborative Research Fund. Y. W. was supported by the National Key R$\&$D Program of China (Grants No. 2019YFA0308000, No. 2020YFA0308800), the
National Natural Science Foundation of China (Grants No. 61971035, No. 61901038, No. 61725107) and Beijing
Natural Science Foundation (No. Z190006, No. 4192054). J. L. was supported by the National Natural Science
Foundation of China (Grant No. 11874036) and Local Innovative and Research Teams Project of Guangdong Pearl River Talents Program (2017BT01N111). H. Y. was supported in part by the National Natural Science Foundation of China (Grant No. 11825404).

\end{acknowledgments}

\end{document}